\pdfoutput=1

\documentclass[superscriptaddress,amsmath,amssymb,twocolumn]{revtex4-1}

\usepackage{graphicx}
\usepackage{chemarr}
\usepackage{times,longtable}
\usepackage{hyperref}
\usepackage{color}
\usepackage{afterpage}
\usepackage{soul,comment}
\usepackage[usenames,dvipsnames]{xcolor}
\usepackage{colortbl}
\usepackage{pbox}
\usepackage{capt-of}% or use the larger `caption` package

\graphicspath{{large/}}

\bibpunct{[}{]}{,}{n}{}{,}

\newcommand{\avg}[1]{\left\langle #1 \right\rangle}

\usepackage{chemarr}
\usepackage{epsf,mathtools}
\usepackage{graphicx}% Include figure files
\usepackage{dcolumn}% Align table columns on decimal point
\usepackage{bm}% bold math
\begin{document}

\title{\textsf{Constraint-based inverse modeling of metabolic networks: a proof of concept%Inferring the probability distribution of reaction fluxes in metabolic networks from empirical flux data: a proof of concept
}}

\author{Daniele De Martino}
\affiliation{Institute of Science and Technology Austria (IST Austria), Klosterneuburg, Austria}

\author{Andrea De Martino}
\affiliation{Soft and Living Matter Lab, Institute of Nanotechnology (CNR-NANOTEC), Consiglio Nazionale delle Ricerche, Rome, Italy}
\affiliation{Human Genetics Foundation, Turin, Italy}

\begin{abstract}
We consider the problem of inferring the  probability distribution of flux configurations in metabolic network models from empirical flux data. For the simple case in which experimental averages are to be retrieved, data are described by a Boltzmann-like distribution ($\propto e^{F/T}$) where $F$ is a linear combination of fluxes and the `temperature' parameter $T\geq 0$ allows for fluctuations. The zero-temperature limit corresponds to a Flux Balance Analysis scenario, where an objective function ($F$) is maximized. As a test, we have inverse modeled, by means of Boltzmann learning, the catabolic core of {\it Escherichia coli} in glucose-limited aerobic stationary growth conditions. Empirical means are best reproduced when $F$ is a simple combination of biomass production and glucose uptake and the  temperature is finite, implying the presence of fluctuations. The scheme presented here has the potential to deliver new quantitative insight on cellular metabolism. Our implementation is however computationally intensive, and highlights the major role that effective algorithms to sample the high-dimensional solution space of metabolic networks can play in this field.
\end{abstract}

\maketitle

Building a system-level understanding of metabolism, the highly conserved set of chemical processes devoted to energy transduction, growth and maintanance in living cells, is a major  challenge for systems biology. Constraint-based {\it in silico} methods like Flux Balance Analysis (FBA) and its refinements play the central role in this endeavour \cite{cbms}. Such schemes provide a coherent theoretical framework to analyze the capabilities of large (genome-scale) metabolic networks starting from minimal genomic input and physico-chemical constraints. Optimal flux configurations are usually hypothesized to optimize a function of the fluxes, like biomass production, leading to results that can be quantitatively tested against experimental data obtained in controlled conditions \cite{compa,bay,biomass}. On the other hand, the choice of an objective function is in many situations not straightforward, and living cells often appear to be multi-objective (Pareto) optimal \cite{sci,hart,mori}. In addition, while bulk properties can usually be well described by optimal flux patterns, experiments performed at single-cell resolution display a considerable degree of cell-to-cell variability \cite{elf}, which has been linked to the inevitable presence of randomness in metabolic processes \cite{tans}. At the simplest level, this suggests that flux patterns observed empirically can be thought to be sampled from a stationary probability distribution defined over the space of feasible metabolic states. This idea is validated by the fact that empirical growth rate distributions are reproduced by a Maximum Entropy principle at fixed average growth rate, according to which flux configurations $\mathbf{v}$ occur with a Boltzmann-like probability distribution 
\begin{equation}\label{unoo}
P(\mathbf{v})\propto e^{\beta \lambda(\mathbf{v})}~~,
\end{equation}
with $\lambda(\mathbf{v})$ the growth rate and where $\beta>0$ controls the magnitude of fluctuations \cite{physbio}. While more comprehensive studies will hopefully refine this picture, recent work has provided further support to and insight into the MaxEnt scenario \cite{ddm}. 

A tightly related question is whether one can infer the probability distribution of metabolic states from empirical data rather than postulating it. Inverse problems and related techniques have a long history in applied fields \cite{tarantola} and have attracted much attention from more theoretical areas in recent years as they directly connect machine learning, artificial neural networks and statistical mechanics \cite{iip}. In the biological context, they have lead to new insights in domains as apart as protein science \cite{cocco}, neural dynamics \cite{front} and immunology \cite{immune}. In this short note we discuss a method to obtain the probability distribution of fluxes in a metabolic network from the experimental characterization of a subset of fluxes, and present a proof-of-concept validation based on flux data for {\it Escherichia coli} steady state growth. Taking experimental means as the key features to be captured, we look for flux distribution of the form
\begin{equation}\label{nostra}
P(\mathbf{v})\propto e^{F(\mathbf{v})/T}~~,
\end{equation}
where $F$ is a function of fluxes (to be inferred) and $T\geq 0$ is an adjustable parameter. In order to learn $F$, we combine a Boltzmann learning scheme with a Monte Carlo sampling method. The  $F$ thus obtained is found to involve a linear combination of the biomass output and of the glucose intake rate, and data are best described by setting $T$ to a finite (non-zero) value, suggesting, in agreement with previous studies, that empirical fluctuations reflect at least in part some unavoidable (and possibly functionally relevant) noise in the organization of flux configurations.

%%%%%%%%%%%%%%%%%%%%%%
%%%%%%%%%%%%%%%%%%%%%%
%%%%%%%%%%%%%%%%%%%%%%

We shall denote by $K$ the number of different samples in which a subset $\mathcal{X}$ of fluxes has been experimentally quantified, and by $v_j^{(k)}$ the value of flux $v_j$ in the $k$-th sample. The empirical mean of flux $v_j$ is given by
\begin{equation}\label{xpmeans}
\avg{v_j}_{{\rm emp}}=\frac{1}{K}\sum_{k=1}^K v_j^{(k)}~~~~~(j\in\mathcal{X}).
\end{equation}
To define the space of {\it a priori} feasible flux configurations for the entire metabolic network, we shall follow the standard route of assuming that viable flux vectors $\mathbf{v}$ are non-equilibrium steady states of the underlying system of reactions. The feasible space $\mathcal{F}$ then corresponds to the solutions of $\mathbf{Sv=0}$, where $\mathbf{S}$ stands for the $M\times N$ stoichiometric matrix ($M$ denoting the number of chemical species, and $N$ that of reactions) and where each flux $v_i$ is constrained to lie within an interval $[v_i^{\min},v_i^{\max}]$, whose bounds encode for the physiologically relevant regulatory, kinetic and thermodynamic constraints. Geometrically, such an $\mathcal{F}$ is a convex polytope. In principle, every point $\mathbf{v}\in\mathcal{F}$ is a feasible flux configuration and configurations are assumed to be {\it a priori} equiprobable. However, the empirical means (\ref{xpmeans}) represent information that can refine this assumption. In particular, one may expect that flux vectors should occur with a probability distribution $P(\mathbf{v})$ such that
\begin{equation}\label{mean}
\int_\mathcal{F} v_j P(\mathbf{v})d\mathbf{v}=\avg{v_j}_{{\rm emp}}~~~~~(j\in\mathcal{X})~~.
%=\frac{1}{K}\sum_{k=1}^K v_i^{(k)}~~,
\end{equation}
Following the Maximum Entropy idea, we focus on the least constrained distribution satisfying (\ref{mean}), which maximizes the entropy 
\begin{equation}
S[P]=-\int_\mathcal{F} P(\mathbf{v})\log P(\mathbf{v})d\mathbf{v}
\end{equation}
subject to (\ref{mean}). This is given by
\begin{equation}\label{maxent}
P(\mathbf{v})\equiv P(\mathbf{v}|\mathbf{c})=\frac{e^{\sum_{j\in\mathcal{X}} c_j v_j}}{Z(\mathbf{c})}~~~~~(\mathbf{v}\in\mathcal{F})~~,
\end{equation}
where $\mathbf{c}=\{c_j\}_{j\in\mathcal{X}}$ denotes the vector of Lagrange multipliers enforcing (\ref{mean}) for each $v_j$, while
\begin{equation}
Z(\mathbf{c})=\int_\mathcal{F} e^{\sum_{j\in\mathcal{X}} c_j v_j}\,d\mathbf{v}
\end{equation}
ensures proper normalization. 

A key question at this point concerns the values of the constants $c_j$. More precisely, can we set them so as to reproduce empirical means most accurately via (\ref{mean})? By straightforwardly maximizing the log-likelihood of the parameters given the empirical data, i.e. \begin{equation}
\mathcal{L}(\mathbf{c}|\mathrm{data})=\frac{1}{K}\sum_{k=1}^K \log P(\mathrm{data}|\mathbf{c})~~,
\end{equation}
one sees that
\begin{gather}
%\mathcal{L}(\mathbf{c}|\mathrm{data})=\sum_i c_i\avg{v_i}_{{\rm emp}}-\log Z(\mathbf{c})~~\\
\frac{\partial\mathcal{L}}{\partial c_j}=\avg{v_j}_{{\rm emp}}-\avg{v_j}_{\mathbf{c}}~~,
\end{gather}
where
\begin{equation}\label{meanc}
\avg{v_j}_{\mathbf{c}}=\int_\mathcal{F} v_j\,P(\mathbf{v}|\mathbf{c})\, d\mathbf{v}~~.
%\avg{v_j}_{\mathbf{c}}=\frac{1}{Z(\mathbf{c})}\int_\mathcal{F} v_j\, e^{\sum_{j\in\mathcal{X}} c_j v_j}\, d\mathbf{v}~~.
\end{equation}
This suggests that the optimal vector $\mathbf{c}$ can be found by an updating dynamics driven by the difference between the empirical mean and the mean computed using the current vector $\mathbf{c}$, i.e. via a Boltzmann learning such as 
\begin{equation}\label{BL}
c_j(\tau+\delta\tau)-c_j(\tau)=\left[\avg{v_j}_{{\rm emp}}-\avg{v_j}_{\mathbf{c}(\tau)}\right]\delta\tau~~.
\end{equation} 

Ideally, the vector $\mathbf{c}^\star$ obtained as the asymptotic fixed point of (\ref{BL}) ensures the best agreement between empirical and theoretical means (i.e. between (\ref{xpmeans}) and (\ref{mean})), while  $P(\mathbf{v}|\mathbf{c}^\star)$ provides our best guess for the (stationary) probability distribution compatible with empirical means that has generated our dataset. In this scenario, the quantity $E=\sum_{j\in\mathcal{X}}c_j^\star v_j$ would represent the key physical parameter regulating the probability of occurrence of feasible flux vectors. Note that such an $E$ plays the role of $F/T$ in (\ref{nostra}).

We have implemented the above scheme using data retrieved from \cite{zhang2014cecafdb}, where {\it E. coli}'s central carbon metabolism is characterized in minimal glucose-limited aerobic conditions, at growth/dilution rates below $0.5/$h. We collected $K=35$ control experiments and the corresponding values for $|\mathcal{X}|=24$ reactions fluxes (see \cite{ddm} for more details). We have then studied the inference problem on the feasible space $\mathcal{F}$ defined by the {\it E. coli}'s core metabolic network \cite{core}. The dimension of $\mathcal{F}$ when a minimal aerobic glucose-limited medium is used to constrain the exchanges between the cell and its surroundings is ${\rm dim}(\mathcal{F})=23$.

To compute the optimal values of the coefficients $c_j$ from (\ref{BL}) we have used the following procedure:
\begin{enumerate}
\item initialize $c_j(0)=0$ for all $j\in\mathcal{X}$
\item at each time step $\tau$: compute $\avg{v_j}_{\mathbf{c}(\tau)}$ from (\ref{meanc}) by sampling the distribution $P(\mathbf{v}|\mathbf{c}(\tau))$ via Hit and Run Monte Carlo \cite{mc}; then
\item find the $j\in\mathcal{X}$ for which the difference $\avg{v_j}_{{\rm emp}}-\avg{v_j}_{\mathbf{c}(\tau)}$ is largest,  update its value according to (\ref{BL}), and iterate.
\end{enumerate}
Finally, we set $\delta \tau =10^{-3}$. 

\begin{figure}
\begin{center}
\includegraphics[width=6.5cm]{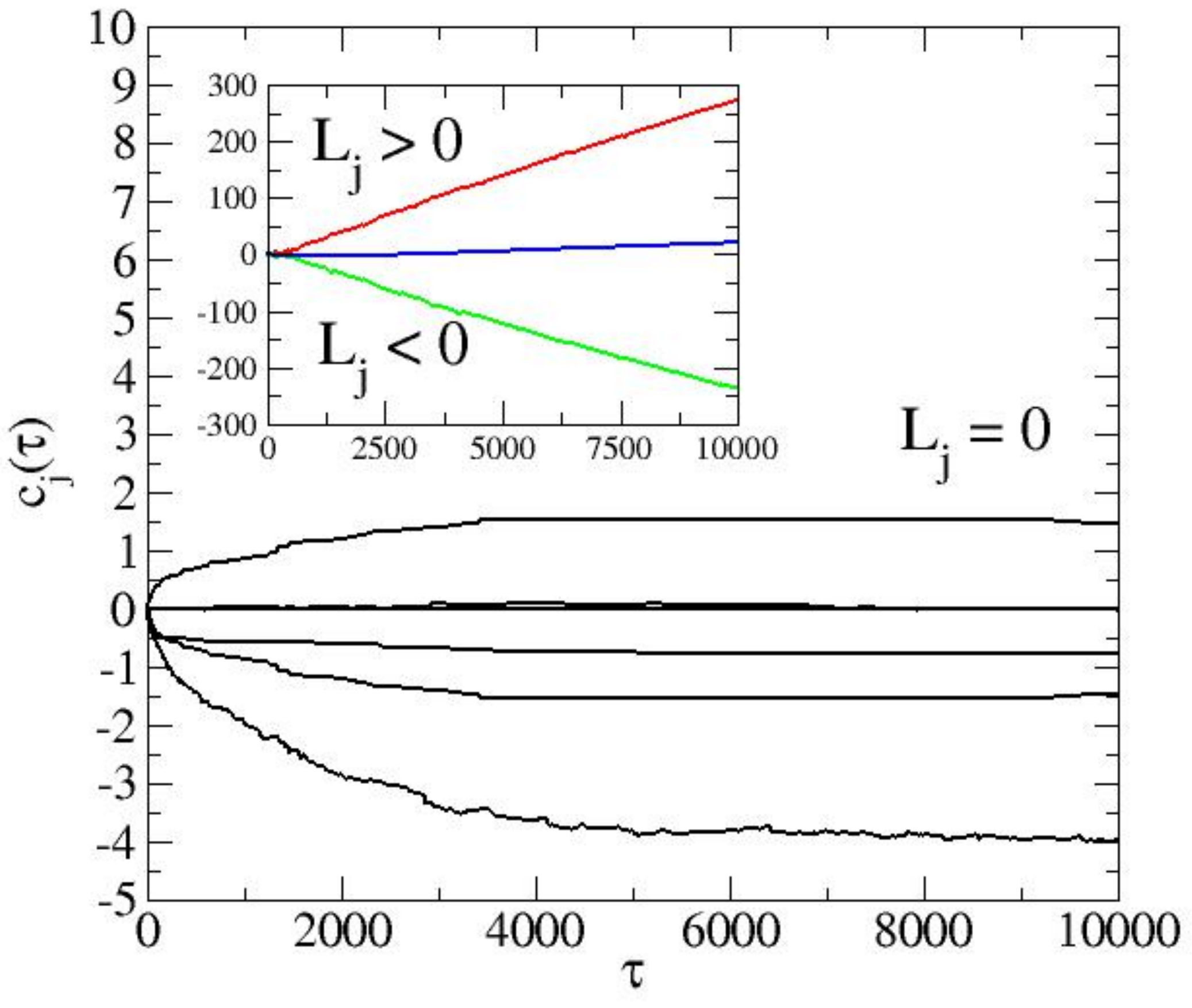}
\end{center}
\caption{Values of $c_j$ as a function of the Boltzmann learning time $\tau$ for the fluxes $j\in\mathcal{X}$. Only a representative sample of the components with $L_j\simeq 0$ is shown, while diverging components are displayed in the inset. \label{one}}
\end{figure}

By studying numerically the dynamics of the $c_j$'s one finds that, at sufficiently long times $\tau$, coefficients generically behave as
\begin{equation}\label{diverge}
c_j(\tau)\simeq L_j \tau~~~~~(j\in\mathcal{X})~~,
\end{equation}
where $L_j$ are flux-specific constants. For most fluxes, though, $L_j=0$, i.e. the corresponding $c_j$'s converge in time to finite (small) values, while $L_j\neq 0$ for a small number of fluxes   (see Fig. \ref{one}). 
This in turn implies that the function
\begin{equation}\label{energy2}
E\simeq \tau\sum_{j\in\mathcal{X}} L_j v_j
\end{equation}
is asymptotically dominated by terms with $L_j\neq 0$. Exploiting the linear dependencies between variables, one can express $E$ in terms of biologically significant fluxes other than those in $\mathcal{X}$. Quite remarkably, one finds that the dominant contribution to $E$ has the form
\begin{equation}\label{energy}
E\simeq \tau (L_\lambda \lambda+L_u u)~~,
\end{equation}
where $\lambda$ and $u$ denote, respectively, the biomass output rate and the glucose in-take flux while $L_\lambda$ and $L_u$ are numerical coefficients given by approximately by $L_\lambda \simeq 2/3$ and $L_u \simeq 1/3$.  

The potentially high-dimensional function (\ref{energy2}) therefore reduces, after Boltzmann learning, to a simple form that combines two  variables of the highest biological significance. In particular, the solution of the inverse problem suggests a scenario very similar to that derived in \cite{physbio}. In addition, though, the network's state also appears to be sensitive to the glucose import rate $u$. 

\begin{figure}
\begin{center}
\includegraphics[width=8.5cm]{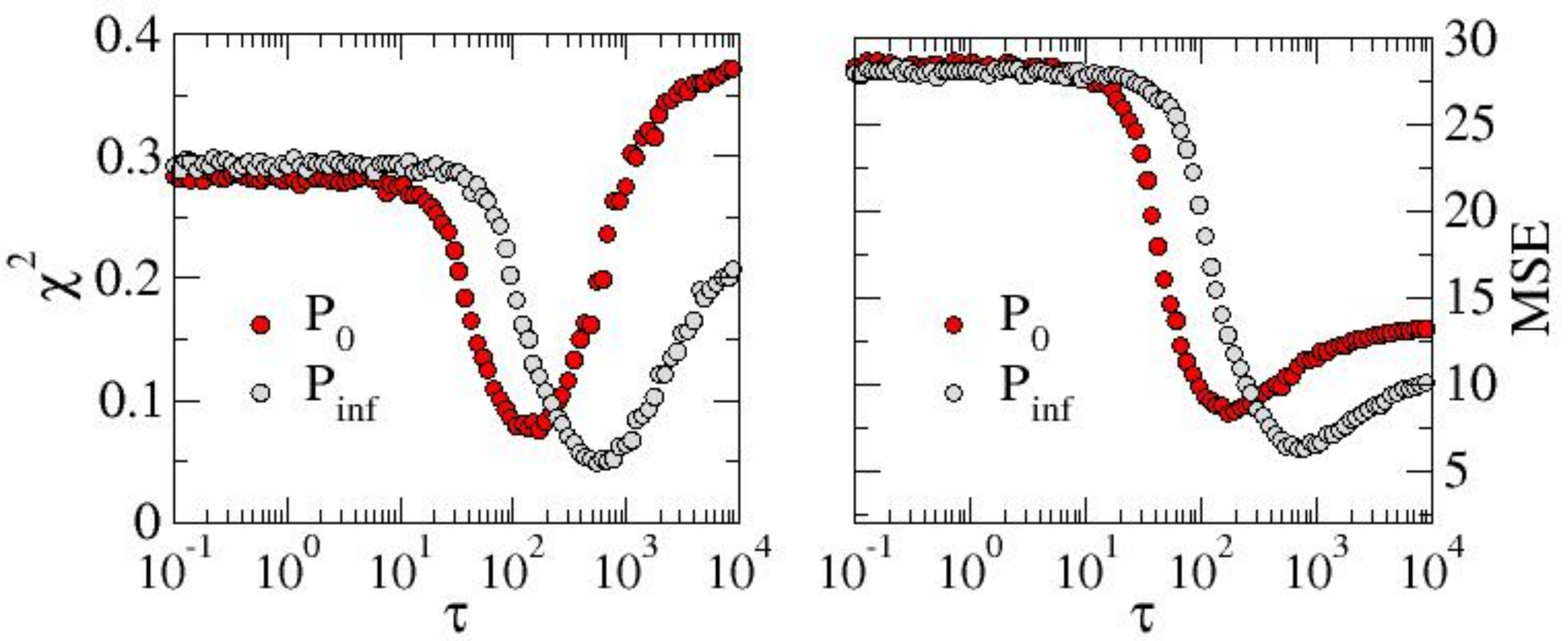}
\end{center}
\caption{Values of $\chi^2$ (left) and MSE (right) as a function of the Boltzmann learning time $\tau$ obtained by comparing experimental data with the inferred distribution (\ref{inf}) (grey markers) and the guessed one (\ref{ninf}) (red markers). \label{two}}
\end{figure}

\begin{figure}
\begin{center}
\includegraphics[width=8cm]{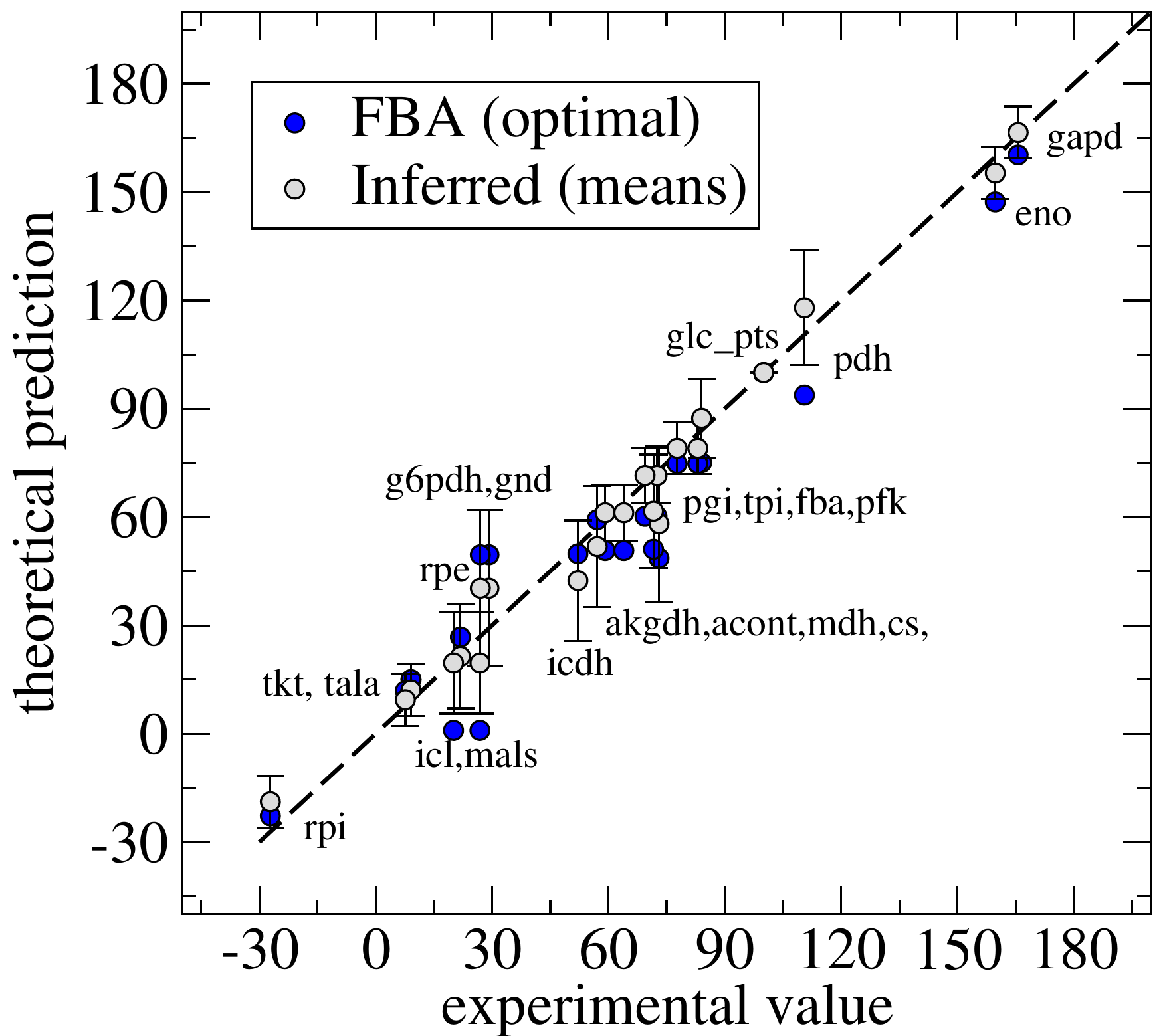}
\end{center}
\caption{Flux-by-flux comparison between experimental values and theoretical predictions obtained from FBA (blue markers) and from the inferred distribution (grey markers). \label{three} Flux intensities are in percentage with respect to the glucose uptake.  Reaction acronyms are taken from \cite{core}.}
\end{figure}
We have validated these results by comparing empirical averages against the means obtained from the inferred distribution
\begin{equation}\label{inf}
P_{\rm inf}(\mathbf{v})\propto e^{\tau (L_\lambda \lambda+L_u u)}~~,
\end{equation}
as well as against means computed from the simpler form (\ref{unoo}), i.e. 
\begin{equation}\label{ninf}
P_0(\mathbf{v})\propto e^{\tau \lambda}~~,
\end{equation}
which was obtained by focusing on growth rate distributions (rather than individual fluxes). In both cases, we have performed an asymptotic extrapolation by studying how $\chi^2$ and the mean squared error (MSE) between data and models change upon varying $\tau$, which --comparing (\ref{nostra}) and (\ref{inf})-- thus plays the role of $1/T$ (while $L_\lambda \lambda+L_u u$ plays the role of $F$). Results are shown in Fig. \ref{two}. One sees that both probability distributions generate clear minima in $\chi^2$ and MSE as functions of $\tau$, where data are optimally reproduced. However, the inferred distribution (\ref{inf}) outperforms (\ref{ninf}) (deeper minimum, albeit slightly) both in terms of $\chi^2$ and in terms of MSE. Interestingly, the inferred function provides better results at a larger value of $\tau$ compared to $P_0$, which suggests that metabolic configurations are closer to optima of $F=L_\lambda\lambda+L_u u$ than to optima of $\lambda$.  In order to get a more precise idea of the improvement obtained via inference, we have displayed in Fig. \ref{three} a detailed flux-by-flux comparison between inferred means (with errors) and experimental means, also showing for sakes of completeness the predictions obtained from standard biomass-maximizing FBA. 

It is important to note that the Boltzmann learning dynamics does not converge in general to a vector $\mathbf{c}$ such that theoretical means match empirical ones perfectly as $\tau\to\infty$. A possible reason is that empirical means lie outside the feasible space $\mathcal{F}$. Flux values are in fact tightly constrained by the mass balance equations $\mathbf{Sv=0}$, and small numerical errors that may arise e.g. from experimental noise can lead to violations of these conditions. While not surprising {\it per se}, this represents a further complication for the inference problem, which for metabolic networks relies on the definition of $\mathcal{F}$, and  further  investigations are needed.

To summarize, previous work has shown that maximum entropy distributions at fixed average growth rate in the space of feasible metabolic states reproduce empirical growth rate distributions measured in exponentially growing populations at single-cell resolution \cite{physbio} and outperform standard FBA (retrieved in the limit $\beta \to \infty$) in reproducing experimental data on fluxes \cite{ddm}. The approach discussed here generalizes the above results by extending the input dataset to a subset of metabolic fluxes. Following standard inference schemes, we have computed the function $F(\mathbf{v})$ that best describes the flux dataset as being generated from a Boltzmann-like distribution $\propto e^{F(\mathbf{v})/T}$. Quite remarkably, the biomass output and the glucose intake emerge from the inference process as the key fluxes that govern the statistics of fluxes. The type of inverse modeling discussed here represents a novel and potentially powerful tool to analyze metabolic networks and characterize their large-scale organization in a way that is fully data-driven. `Energy' functions $F$ obtained in this way may provide a new perspective on metabolic functions and objectives in complex settings where biomass growth optimization alone does not suffice to explain observations. However, the predictive capacity of inverse methods is intrinsically limited by the quality of available data. In addition, the scheme employed here has a high computational cost. In fact, as a Monte Carlo sampling of the feasible space $\mathcal{F}$ is required at every Boltzmann learning step, performing the same analysis on larger (genome-scale) networks, with a feasible space $\mathcal{F}$ whose dimension can be an order of magnitude larger than that addressed here, is very difficult. This problem may however be effectively solved by the use of approximate representations of the feasible space and/or of more efficient computational heuristics \cite{alfre, cossio}. In this sense, the present note represents merely a proof of concept and more work is needed to make these ideas viable at genome resolution.

We thank A. Braunstein, A. P. Muntoni and A. Pagnani for a critical revision of these ideas and for important comments and suggestions. We acknowledge the support of the People Programme (Marie Curie Actions) of the European Union's Seventh Framework Programme (FP7/2007-2013) under REA grant agreement no. [291734] (D.D.M).

% https://academic.oup.com/bioinformatics/article/23/3/351/236644/Bayesian-based-selection-of-metabolic-objective

\end{document}